\documentclass[12pt]{iopart}

\usepackage{graphicx,amssymb,ulem}   %amsmath,
\usepackage{textcomp}
\usepackage{xcolor}
\usepackage{hyperref}
\usepackage{color}
\usepackage{epstopdf}
\usepackage{nicefrac}
\usepackage{gensymb}
\usepackage{placeins}
\usepackage{xspace}
\usepackage{lineno}
\usepackage[hypcap]{caption}
\usepackage{appendix}

\def\40K{$^{40}$K}

\def\X{$X^2\Sigma^+\,$}
\def\B{$B^2\Sigma^+\,$}
\def\A{$A^2\Pi_{1/2}\,$}


\begin{document}

\title{Maximizing the capture velocity of molecular magneto-optical traps with Bayesian optimization} 

\hspace*{\fill} \\\noindent\textbf{Supeng Xu, P Kaebert, M Stepanova, T Poll, M Siercke$^1$, and S Ospelkaus$^1$}

 \noindent {Institut f{\"u}r Quantenoptik, Leibniz Universit{\"a}t Hannover, 30167 Hannover, Germany}
 $^1$To whom any correspondence should be addressed        

\noindent\textbf{Email:} siercke@iqo.uni-hannover.de \textbf{and} silke.ospelkaus@iqo.uni-hannover.de

\date{\today}

\begin{abstract}
 Magneto-optical trapping (MOT) is a key technique on the route towards ultracold molecular ensembles. However, the realization and optimization of magneto-optical traps with their wide parameter space is particularly difficult. Here, we present a very general method for the optimization of molecular magneto-optical trap operation by means of Bayesian optimization (BO). As an example for a possible application, we consider the optimization of a calcium fluoride (CaF) MOT for maximum capture velocity. We find that both the \X to \A and the  \X to \B transition  to  allow for capture velocities larger than  $20$~m/s with $24$~m/s and $23$~m/s respectively at a total laser power of $200$\,mW. In our simulation, the optimized capture velocity depends logarithmically on the beam power  within the simulated power range of  $25$ to $400$\,mW.  Applied to heavy molecules such as BaH and BaF with their low capture velocity MOTs it might offer a route to far more robust magneto-optical trapping. 
\end{abstract}

\maketitle
\section{Introduction}

Cooling molecules to temperatures near absolute zero is expected to open many exciting research opportunities ranging from dipolar quantum many-body physics \cite{cite1} to precision tests of fundamental physics \cite{cite2,cite3,cite4}. Direct laser cooling and trapping \cite{cite5}, which is the workhorse of cold atom physics, have already been successfully developed for molecules. While tremendous progress has been achieved in laser cooling of molecules \cite{cite6,cite7,cite8}, magneto-optical trapping \cite{cite9,cite10,cite11,cite12}, and conservative trapping using magnetic \cite{cite13,cite14} and optical \cite{cite15} traps, the number of molecules loaded from the molecular beam is currently still limited to about $10^5$ \cite{cite11,cite16}. The biggest limiting factors are the low efficiency of current slowing methods \cite{cite17,cite18} along with the generally lower flux beam source \cite{cite19,cite20}. Some methods have been proposed to increase the number of slowed molecules, such as the Zeeman–Sisyphus decelerator \cite{cite21} and the molecular Zeeman slower \cite{cite22,cite23}, or to increase the loading efficiency using the stimulated emission force \cite{cite24} , but this increase in efficiency still remains to be proven experimentally. 
\par Zeeman slowing combines the advantages of white light slowing and chirped-light slowing as it is a continuous method and provides molecules with a well-defined final velocity \cite{cite25}. However, after passing the slowing region, the molecules will have a period of free flight before they reach the center of the trap. This can be problematic for MOTs that have low capture velocities and rely on a maxed-out slowing region. Since the remaining transverse velocity of the molecules will result in the expansion of the molecular cloud, a low longitudinal velocity at the end of the slower means a longer time of flight, and a larger molecular cloud reaching the trap center. Transverse cooling of the molecular beam ahead of the slowing process can have an effect on increasing the number of loaded molecules, but the improvement is limited \cite{cite26}, since the initial collimated beam will finally expand again due to the directional absorption and random re-emission of photons \cite{cite27}. In contrast, an improvement to the capture velocity of the MOT will increase the MOT population by orders of magnitude \cite{cite28}, which is what we are seeking to do in this manuscript. 
\par Previous theoretical studies of molecular MOTs always kept a common detuning of each laser component to its respective transition, evenly distributed laser intensity among each laser component, and fixed magnetic gradient and beam diameter \cite{cite29,cite30,cite31} to search for the optimal laser polarization configuration. This is reasonable since it is hard to select the optimal configurations from such a high-dimensional parameter space. 
\par In this paper, we use Bayesian optimization \cite{cite32} together with a rate equation model \cite{cite33} to globally search the total parameter space (detuning, polarization and intensity of each laser frequency component, separately, combined with $1/e^2$ beam radius $w$ and magnetic field gradient A) to find the MOT configurations that have the largest capture velocity for both the \X to \A and \X to \B transition.  After  initial selection, we comprehensively consider five aspects of the force profile, including the trap frequency, the damping coefficient, the peak values of both trapping and cooling acceleration and the capture velocity, to choose the optimal configurations for the MOT. The optimization tells us that the  ``dual-frequency'' effect is still the main reason  responsible for the force of the $X \rightarrow A$ DC MOT, but there are experimentally feasible laser configurations that can provide a better confining and damping force than the configurations currently used. For the case of the \X to \B  transition, which is thought to be not suitable for a MOT because of the large energy splitting of the upper F = 0 and F = 1 levels \cite{cite29}, we also find some configurations that provide a large capture velocity and a considerable MOT force. \footnote{Note that a realistic scheme for a \X to \B state MOT would have to consider a leakage to the \A state. However, since the order of magnitude of this leakage is not yet known, this effect is neglected here.} We investigate the dependence of capture velocity on the total laser power and find a good linear fit to the logarithm of laser power from 25 to 400 mW. Our method can be applied to heavy molecules like SrF \cite{cite7}, BaF \cite{cite23} and BaH \cite{cite31} to search for solutions with multiple frequency components and improve the loading efficiency of the MOT.

 \section{Rate equation model}

\begin{figure}
\centering
\includegraphics[scale=0.6]{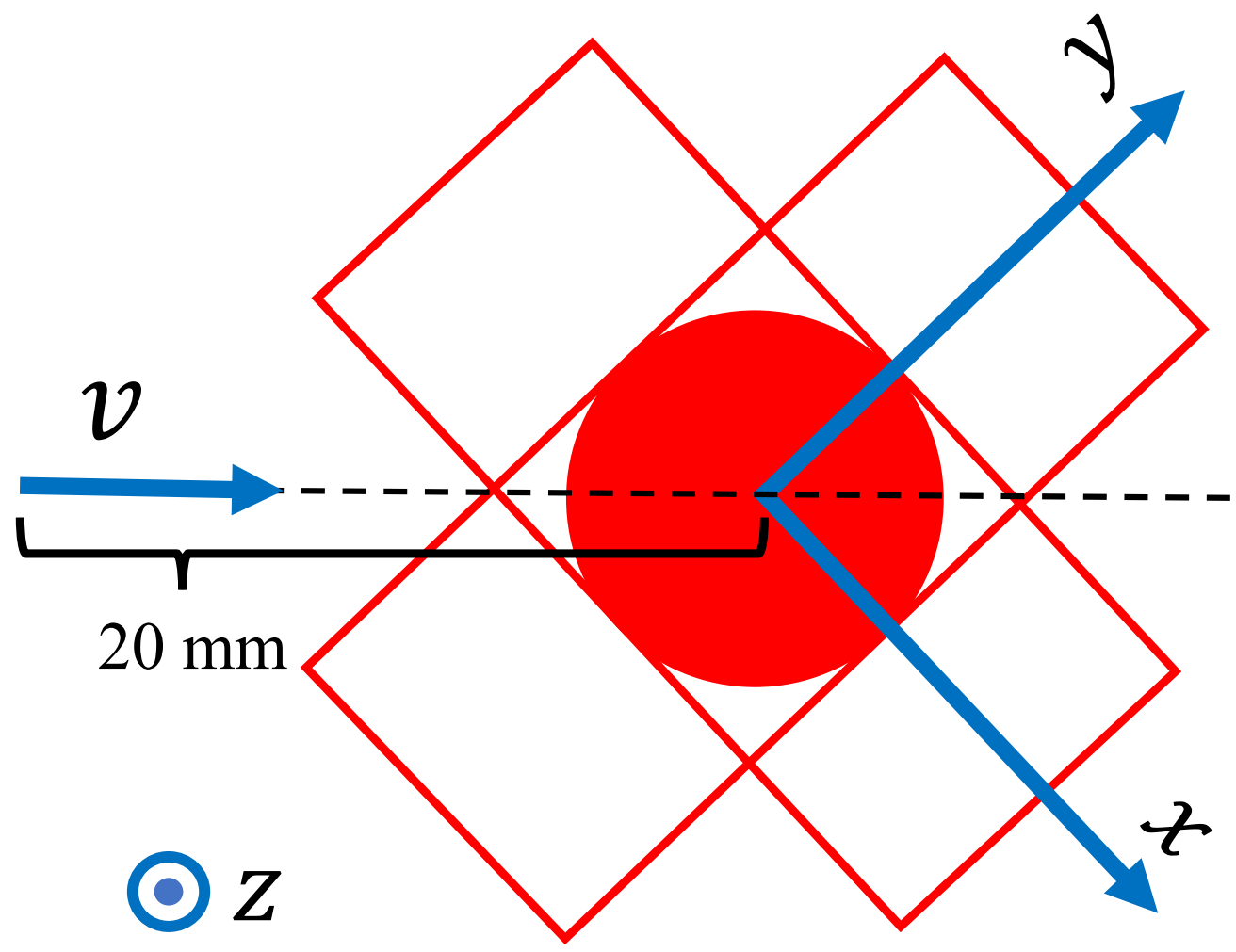}
\caption{Diagram of the  scheme used to simulate the MOT and the capture velocity. The two crossed and perpendicular red hollow rectangles represent the  counter-propagating laser beams in the $xy$ plane, the red filled circle  refers to the third laser beam  pair counter-propagating along  the $z$ axis.}
\label{Fig1}
\end{figure}

 We use the rate equation model described in \cite{cite33} to calculate the force of the CaF MOT, in which we consider the effect of each laser frequency component from all six directions. To find the capture velocity for a specific MOT configuration, we place the molecules at a distance of 20 mm away from the center of the MOT in the $z=0$ plane, traveling along the $\frac{1}{\sqrt{2}}\left(\vec x+\vec y\right)$ direction with a range of speeds (see Fig.\,\ref{Fig1}). The location of the molecule is used to determine the magnetic field according to the formula: $\vec B = A(\vec x + \vec y -  2\vec z)$, which defines the direction of the quantization axis (QA). Then, we use a rotation matrix to project the effect of any laser polarization on the QA 
direction to get the real transition ratio in ($\sigma^+,\sigma^-,\pi$). For details, see \ref{Appdx:Rot_Mat}. The velocities along each axis are used to get the Doppler shift. The validity of our code is verified by reproducing the results in \cite{cite29}. 

\begin{figure}
\centering
\includegraphics[scale=0.5]{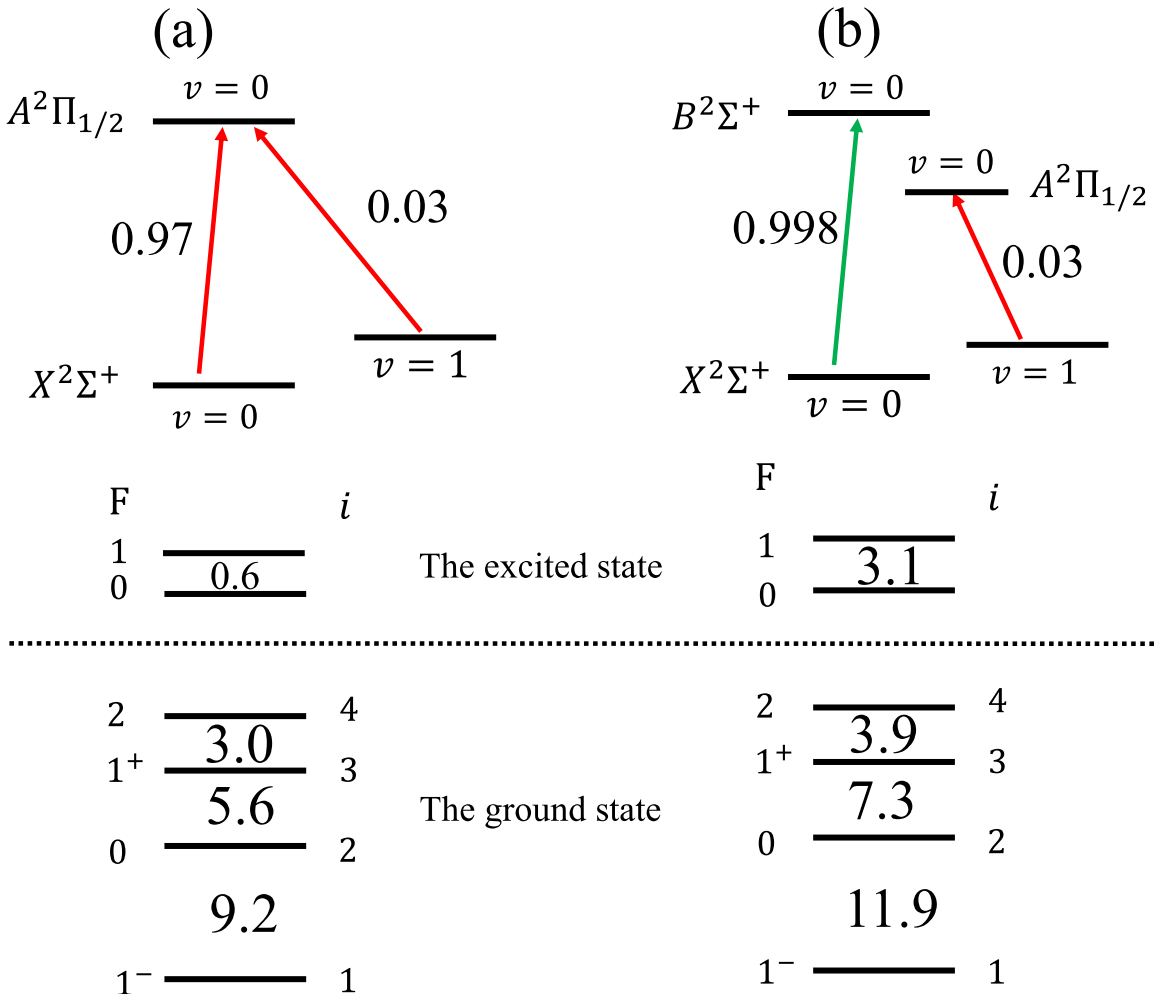}
\caption{CaF MOT transitions used in the simulation. (a) The \X $\rightarrow$ \A  transition. Since both $v=0$ and $v=1$ lasers address the same upper state, both need to be 
considered in the rate equations. The lower part shows the hyperfine states where the energy splitting between interconnected levels are labeled in units of $\Gamma$. (b) 
The \X $\rightarrow$ \B transition. While re-pumping due to vibrational relaxation is necessary for this configuration, just as was the case in the \X $\rightarrow$ \A transition, the $v=1$ state does not need to be considered in the rate equations since it is pumped over a different excited state, thereby having little influence on the final force. The spacings between hyperfine states are labeled in units of the decay rate from the \B state.}
\label{Fig2}
\end{figure}

As is shown in Fig.\,\ref{Fig2}, for the \X to \A transition, both \X (\textit{v} = 0) to \A (\textit{v} = 0) and \X(\textit{v} = 1) to \A(\textit{v} = 0) vibrational transitions are considered to correctly model the excited state populations, while for the \X to \B transition, since the ground $v=0$ state and the $v=1$ state do not share the same excited state and the Franck-Condon factor of the \X(\textit{v} = 0) to \B(\textit{v} = 0) transition is nearly unity, we only consider the main pump laser. We neglect higher vibrational states, because their effect on the size and shape of the MOT force is too small. Through the whole simulation, the fully nonlinear Zeeman splittings of \X $(v=0,1, N=1)$, \A $(v= 0, J = 1/2)$, and \B $(v=0, N=0)$ ro-vibrational states are used. 

The ground and excited state hyperfine levels of both \X $\rightarrow$ \A and \X $\rightarrow$  \B are shown  in the lower part of Fig.\,\ref{Fig2}, along with the energy splitting in units of decay rate $\Gamma$. In the following discussion, the $F = 1^-, 0, 1^+\, \textrm{and\,}2$ are labeled as $i=1, 2, 3, 4$, respectively. In the simulation, each level $i$  is addressed by an individual laser with independent polarization, intensity ratio and detuning, ($\sigma_i, I_i, \delta_i$). In order to catch any influence originating from off-resonant laser excitation, we consider the interactions of each laser with all the four hyperfine levels of the ground state. This approach works quite well for predicting the CaF spectrum in both
low and high magnetic fields\,\cite{cite37}. In order to restrict the calculation to reasonable values and to speed up the process, each parameter is constrained to lie within certain bounds (see Table 1). 

\begin{table}
\begin{center}
\caption{Parameters adjusted by the Bayesian optimization algorithm and their range }
 \begin{tabular}{||c| c ||} 
 \hline
 Variables & Range  \\ [0.5ex] 
 \hline\hline
 $\sigma_i$ & $\sigma^+/\sigma^-$ \\ 
 \hline
 $I_i$ & $[0,1]$ \\
 \hline
 $\delta_i$ & $[-10,10]\Gamma$ \\
 \hline
 $A$ & $[5,30]$\,G/cm \\
 \hline
 $w$ & $[4,15]$\,mm \\ [1ex] 
 \hline
\end{tabular}
\end{center}
\label{tb1}
\end{table}

%(\Gamma$\, to\,\,$)

Note that the detuning is relative to the frequency difference between level $i$ and the upper $F = 1$ state. In the simulation the total laser power is fixed and distributed among the various frequency components according to $I_i$. If there is a fifth frequency component, we choose its detuning such that $\delta_5=0$ when it addresses the $i=2$ level, and allow for $\delta_5=-15\Gamma$ to $15\Gamma$. In conclusion, our model thus ends up with 14 tunable parameters in the case of 4 frequency components, and 17 in the case of 5, a parameter space that would be too large to span by simple loop methods. 

\section{The Bayesian optimization method}

Bayesian optimization is a method particularly suited to the problem at hand. The method requires no knowledge or assumption about the function we are trying to evaluate, and it is especially efficient at finding global maxima \cite{cite34,cite35}. The method models the function we are trying to find as a multivariate Gaussian process \cite{cite36}. The optimization starts off with a prior distribution, and, using a particular acquisition function (e.g. which point has the highest chance of being the maximum), it chooses a point in parameter space to sample. The thereby obtained information is used to update the prior, generating a posterior distribution with less uncertainty around the parameter space where the sample was taken. By treating this posterior as the prior of the next iteration, Bayesian optimization can efficiently sample a parameter space of up to 
20 dimensions to find the global maximum. The global nature of the method is ensured, because areas in parameter space that have not yet been explored carry with them a 
large uncertainty, and will therefore eventually have a higher chance of containing a maximum when compared to areas that have been excessively sampled. On the other 
hand, the method tends to quickly converge to the maximum solution, as it chooses to sample those areas of parameter space that are most likely to contain the maximum. 
\par Our simulation is based on the  \textit{bayesopt} function of Matlab Statistics and Machine Learning Toolbox, which internally uses Gaussian process regression to model the objective function. We use the default acquisition function, \textit{`expected-improvement-per-second-plus'} and a parallel mode to improve the calculation 
speed. As such, even though we don’t know the exact relationship between the laser parameters and the capture velocity, we can calculate the force under a specific laser configuration and 
simulate a MOT with this force to get the maximum capture velocity. %(\textcolor{blue}{\sout{The whole process is a black-box function, which is set to be the surrogate objective function.}})
Note that the \textit{bayesopt} function always searches for the minimum value. To get the maximum value, we therefore turn the capture velocity into a negative number in the program. 
\section{Results}
Using Bayesian optimization we try to find the set of  parameters giving the maximum capture velocity of a MOT for both the \X $\rightarrow$ \A and \X $\rightarrow$  \B transition of CaF. We set a total power of $200$\,mW for the main pump laser, and distribute it according to the number and intensity ratio of laser components. In the cases where there is a re-pump laser, its power is set to be the same as the pump laser but evenly distributed among the four hyperfine states and the values of detuning are all set to zero. The polarization of the re-pump laser has little effect on the final result. In this paper, we therefore set all the polarization from the positive direction to be $\sigma^+$. We find both the \X $\rightarrow$  \A and \X $\rightarrow$  \B transition to have nearly the same maximum capture velocity. We also study the dependence of maximum capture velocity on laser power and observe a logarithmic dependence of the capture velocity on the laser power.  Additionally, we check whether adding an additional frequency component will result in an even better MOT operation, but do not find a significant improvement.  We therefore restrict our discussion to the  four laser frequency case. 

\subsection{The \X to \A transition}

\begin{figure}[htbp] 
\centering
\includegraphics[scale=0.6]{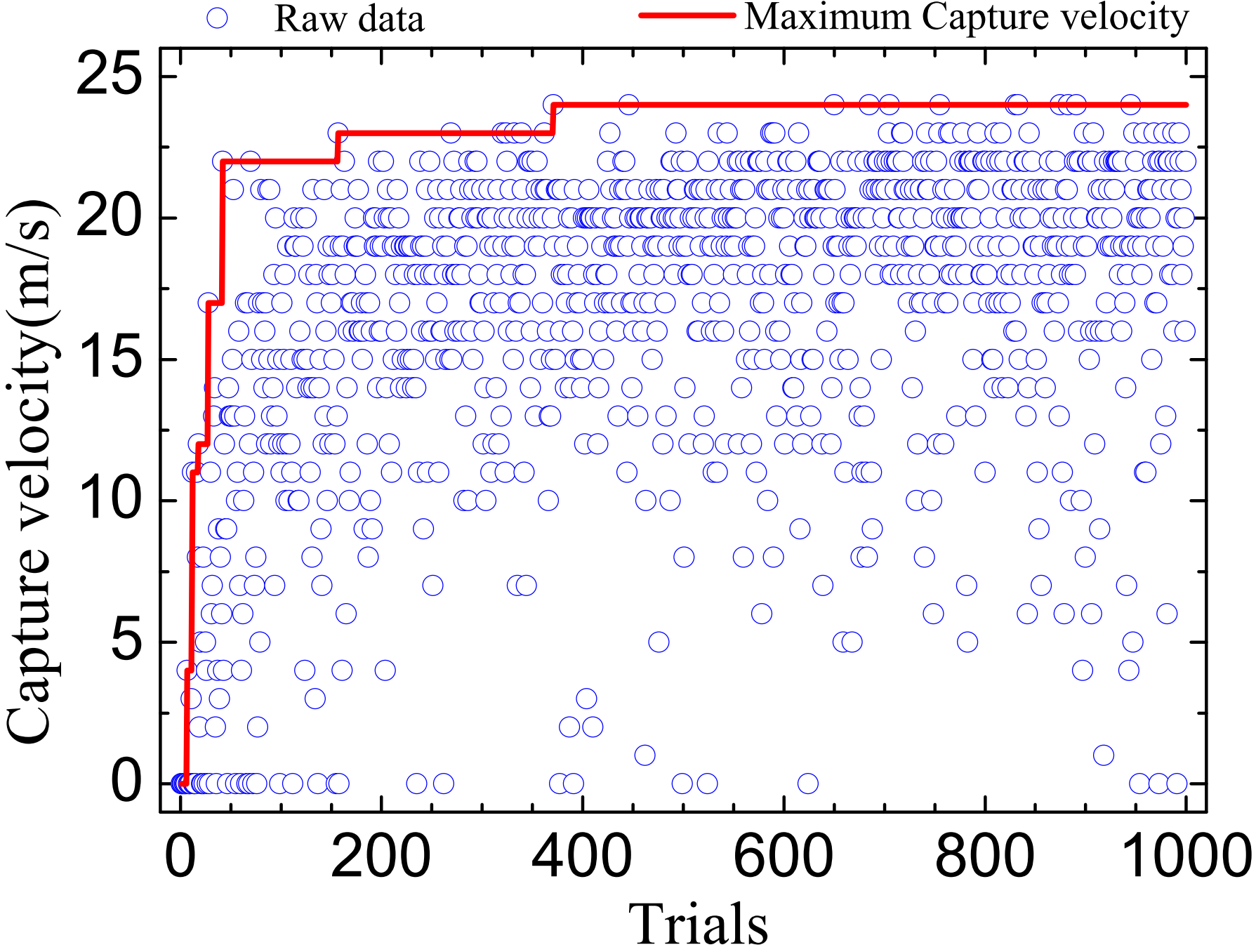}
\caption{Optimization of the capture velocity of an \X $\rightarrow$  \A MOT with a total power of $200$\,mW. The blue circles show the capture velocity for each iteration of the BO
algorithm. The red solid line represents the maximum capture velocity among all the searched results up to that trial. 
}
\label{Fig3}
\end{figure}

The optimization results for the \X to \A transition are shown in Fig.\,\ref{Fig3}. We plot the capture velocity for each iteration of the code.  The red line is the maximum capture velocity observed as a function of iteration number. Starting from 0 m/s, the maximum value of 24 m/s is obtained after $ \sim $400  iterations. We continue the calculation until 1000 iterations to check for even better results. It is worth keeping in mind that, while we will concentrate on the laser configurations for the 24 m/s capture velocities in the following discussion, configurations with slightly lower velocities may be meaningful to consider if other factors (such as MOT size or temperature) are more important, because the program always pursues the maximum capture velocity and ignores the balance of confining and damping force. 

\begin{table}[htbp] 
\centering
\caption{A group of optimized parameters that can give a maximum capture velocity of $24$\,m/s for the \X $\rightarrow$  \A MOT at a power of $200$\,mW. }
\includegraphics[scale=0.22]{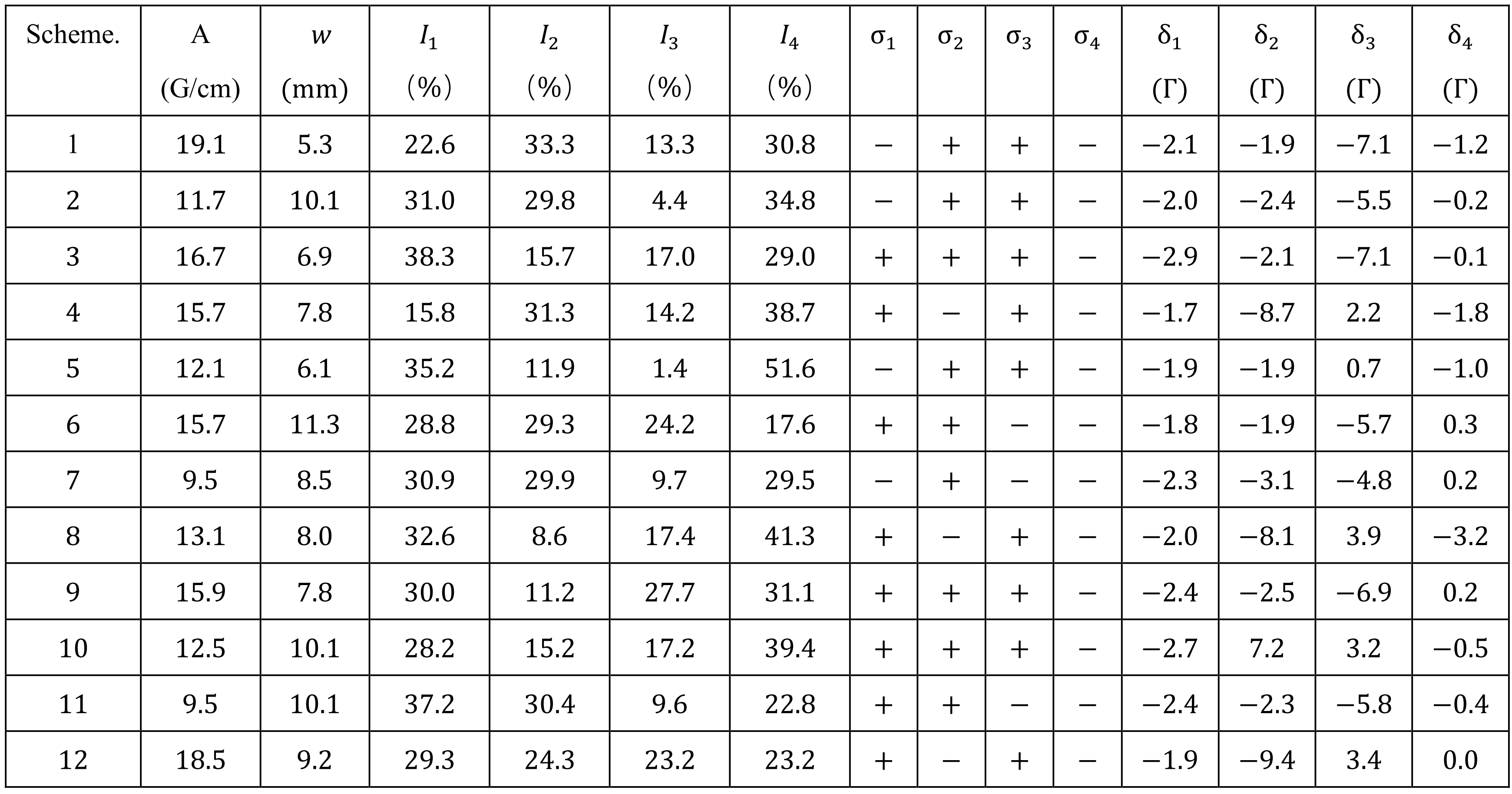}
\label{Table2}
\end{table}

Table 2 lists the set of parameters that can provide capture velocities of $24$\,m/s. All the numbers are rounded to one decimal place. For simplicity, we label a specific set of parameters with scheme index $j $ where $j=1,\, 2,\, \dots,\,\!12$. We calculate both the acceleration of a molecule at rest as a function of its displacement from the center of the trap along $\frac{1}{\sqrt{2}}(\vec x+\vec y)$ direction, and the acceleration of a molecule at the center of the trap versus 
its speed for all the schemes in table 2. We find scheme 1 and scheme 5 have better performance in terms of cooling and trapping force in addition to capture velocity and are fairly straightforward to realize experimentally.
%For the $i=1, 2$ levels, it seems that a detuning of $-2\Gamma$ is beneficial to large capture velocity, while for the $i=4$ level, a detuning of $-\Gamma$ and relatively high proportion of laser %intensity is good for trapping the molecules tightly. The interesting part lies in the laser component addressing the $i= 3$ level, which shows huge changes of where it should be and how much power it %should have. In scheme 1, a detuning of $-7.1\Gamma$ means a detuning of $-4.1\Gamma$ relative to $i= 4$ level, which is still far away from resonance. \textcolor{red}{Without this laser component, the %capture velocity decreases to 23 m/s, but will simplify the experiment. }In scheme 5, a tiny intensity 
%$I_3= 1.4\%$ with positive polarization and $ 0.7\Gamma$ detuning plays a vital role in confining the molecules, as we will discuss later. Note that in \textcolor{red}{scheme 1}, the ``dual-frequency'' %effect is \textcolor{red}{mainly constructed by $I_2$ and $I_4$ laser components}. 
\begin{figure*}
\centering
\includegraphics[scale=0.35]{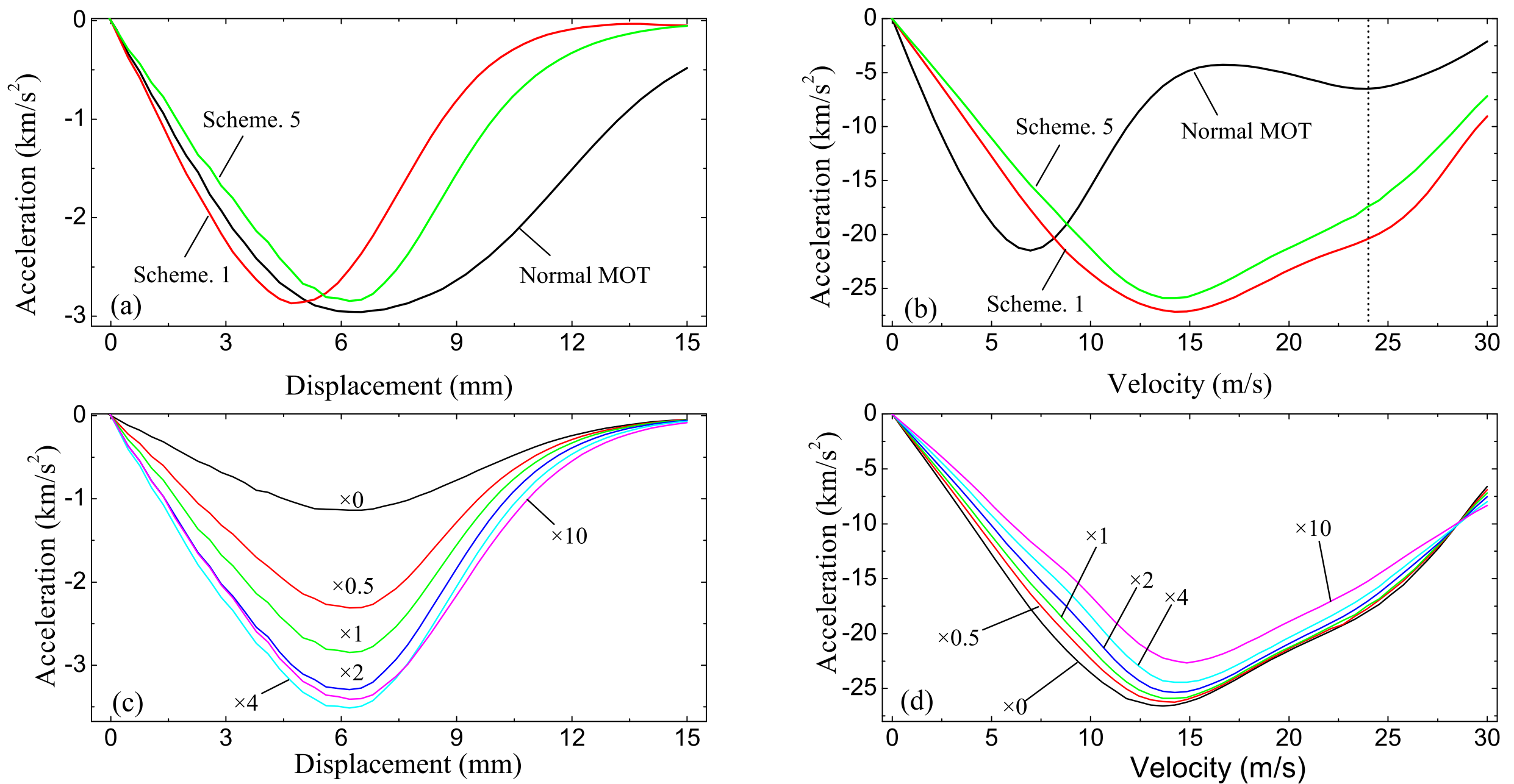}
\caption{Acceleration versus (a) displacement and (b) velocity, for two of the optimized schemes after comprehensively considering trap frequency, damping coefficient and 
peak values of both trapping and cooling force. The dashed vertical line reflects the maximum capture velocity, 24 m/s. The acceleration of the currently employed MOT scheme is also plotted for comparison. The laser power is fixed to $200$\,mW in total. Acceleration versus (c) displacement and (d) velocity, for scheme 5, where $I=1.4\%$ is multiplied by various factors (0, 0.5, 1, 2, 4, 10), keeping the other intensities fixed. 
}
\label{Fig4}
\end{figure*}

\begin{figure}
\centering
\includegraphics[scale=0.55]{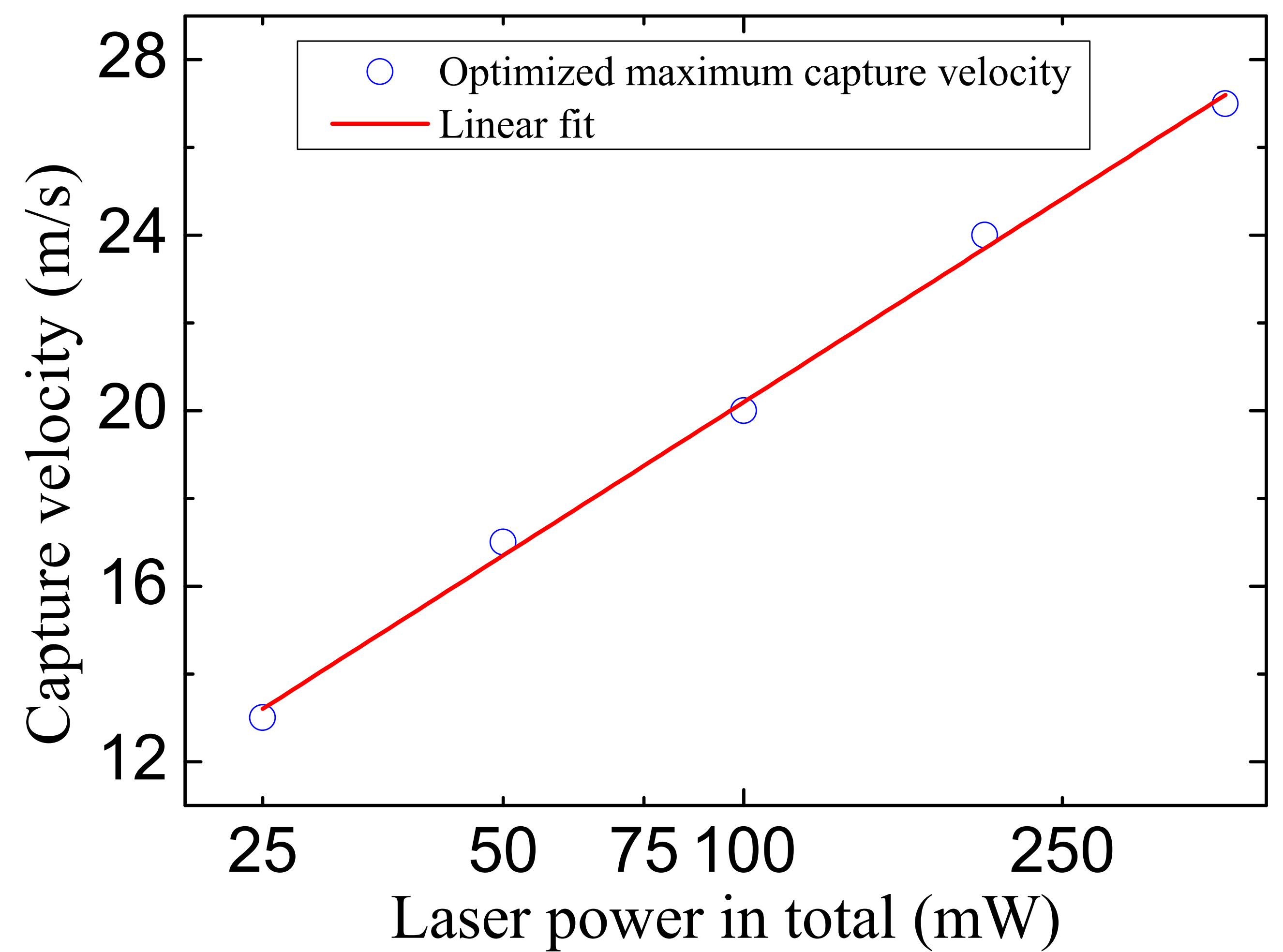}
\caption{Maximum capture velocity versus total laser power for \X $\rightarrow$ \A\,MOT. The $x$ axis is in logarithmic scale.  
}
\label{Fig5}
\end{figure}

The corresponding trapping and cooling acceleration curves are illustrated in Fig.\,\ref{Fig4}(a) and (b), where we compare the results for scheme 1 and 5 to the results of  a conventional  MOT configuration %(\textcolor{red}{suggested optimal polarization configuration in )
with evenly distributed laser power among the four hyperfine states, a common detuning of $-\Gamma$, as well as $A=15$\,G/cm and $w=8\,$mm. %(\textcolor{red}{As can be seen in Fig.\,\ref{Fig4}(a), scheme 1 and scheme 5 have nearly the same peak acceleration as the normal MOT scheme, ~3000 $\ m/s^2$.})
We see (compare Fig. \ref{Fig4}(a) ) that the confining forces for schemes 1 and 5 are similar to the conventional MOT scheme, while the cooling force is significantly extended towards larger velocities for the optimized schemes (compare Fig.\,\ref{Fig4} (b)). %Scheme 1 has a trap frequency of $144$\,Hz, while scheme 5 has a trap frequency of $127$\,Hz, compared with $140$\,Hz for the normal MOT. The vertical dashed line in Fig.\,\ref{Fig4}(b) is the maximum capture velocity position. The normal MOT scheme shows a maximum of the cooling force at ~$7$\,m/s ($= \Gamma/k\times\sqrt{2}$), while scheme 1 and scheme 5 have a peak at $14$\,m/s. The horizontal dashed line in Fig.\,\ref{Fig4}(b) reflects the peak level of cooling acceleration of a normal MOT scheme, corresponding 
%to $22.5$\,m/s of scheme 1 and $19.5$\,m/s of scheme 5. On the other hand, the damping coefficient decreased from $4571$\,s$^{-1}$ of normal MOT scheme to $2530$\,s$^{-1}$ of scheme 1 and $2188$\,s$^{−1}$ of scheme 5. \textcolor{red}{The broad damping force profiles of scheme 1 and scheme 5 ensure the fast deceleration of large velocity molecules while the narrow confining force profile means a relatively smaller final MOT size. }

In scheme 5, it is striking that there is a particular frequency component with a very low intensity fraction ($I_3= 1.4\%$).  It is therefore interesting to investigate whether $I_3$ can be set to zero. The force for various factors of $I_3 = 1.4\,\% \times n$ ($n = 0, 0.5, 1, 2, 4, 10$) is shown in Fig.\,\ref{Fig4}(c) and (d). Unexpectedly, when $n = 0$ , we find the peak value of trapping acceleration is reduced by a factor of ~3, while the cooling acceleration has changed only a little. From n = 0 to n = 4, the trap frequency and the peak trapping acceleration amplitude keep increasing, and stay strong even up to $n = 10$. In contrast, the damping coefficient and the peak cooling acceleration amplitude decrease with the increase of $n$. All these characteristics show that the optimization results are stable and that one can make fine tuning around the optimal configurations to potentially simplify the experiment. 

Lastly we study the dependence of the maximum capture velocity on the total power. For each power, we run the BO algorithm and find the optimized maximum capture velocity. The result is shown in Fig.\,\ref{Fig5}, where the blue circles are simulated results, and the red solid line is a linear fit (note the log scale of power in the plot). Each point is the optimized result after 500 trials. From $25$\,mW to $400$\,mW, the velocity changes from $13$\,m/s to $27$\,m/s.

\subsection{The \X to \B transition}

\begin{figure}
\centering
\includegraphics[scale=0.6]{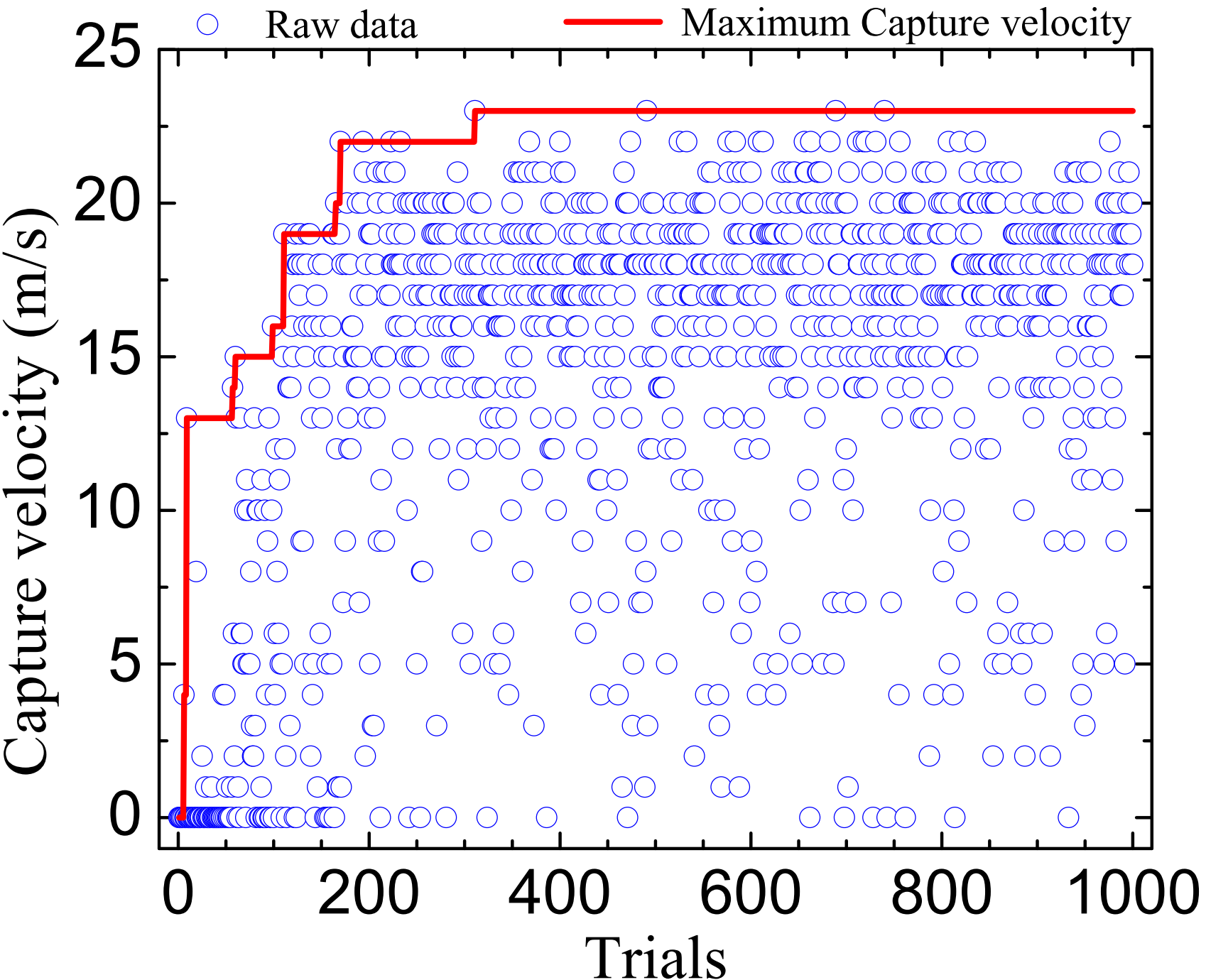}
\caption{Optimization of the capture velocity of an \X $\rightarrow$ \B MOT at a power of $200$\,mW. The blue circles show the capture velocity for each of the 1000 trials of the program. 
The red solid line represents the maximum capture velocity among all the searched results up to that trial. 
}
\label{Fig6}
\end{figure}

The \X to \B  transition has more favorable Franck-Condon factors, and the re-pump lasers don’t need to share the same excited states with the pump laser, which means we can get larger scattering rates while using less power for the re-pumper. However, the hyperfine splitting of $20$\,MHz in the upper (\B) state has made the \B state unsuitable for a MOT with current schemes [29]. While the regular MOT scheme does not seem to work well, using the BO method we find a maximum capture velocity of $23$\,m/s, along with lots of other choices with moderate capture velocities. The optimization process is shown in Fig.\,\ref{Fig6}, where the circular points are various trials, and the red solid line is the 
maximum capture velocity up to that number of iteration. We found four different sets of configuration that provide the maximum capture velocity while still being tractable 
in an experiment. The specific parameters for these four configurations are listed in Table \ref{Table3}. Some interesting observations can be made from the data in the table. First, the closest laser component to the $i = 4$ level is $I_3$, with a $\sigma^-$ polarization, while laser $I_4$, with $\sigma^+$ polarization, mainly addresses the $i = 3$ level. This is totally different from the suggested “dual-frequency” arrangement of the $X \rightarrow\, B$ transition MOT [29], where positive and negative polarization laser components address $i=4$ and $i=3$ states, separately. Furthermore, the nearest laser component to $i=1$ level is $I_2$ in scheme 3, which is still $>5\Gamma$ detuned from the $i=1$ level. This is because of the energy difference of $3.1\Gamma$ in the $B$ state, such that a detuning of $-5.1\Gamma$ relative to the upper $F=1$ level means a detuning of $-2\Gamma$ relative to the upper $F=0$ state, and the $i=1$ level  is  mainly coupled to the upper $F=0$ state. We also notice that the small proportion of laser intensity of $I_4$ is necessary for the large confining force.

\begin{table}[htbp]
\centering
\caption{Experimentally feasible groups of optimized parameters that can give a maximum capture velocity of $23$\,m/s for the \X $\rightarrow$  \B MOT at a power of $200$\,mW. }
\includegraphics[scale=0.75]{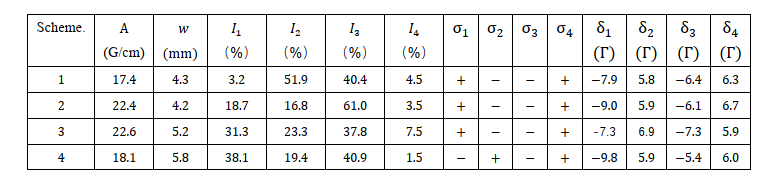}
\label{Table3}
\end{table}

\begin{figure*}
\centering
\includegraphics[scale=0.5]{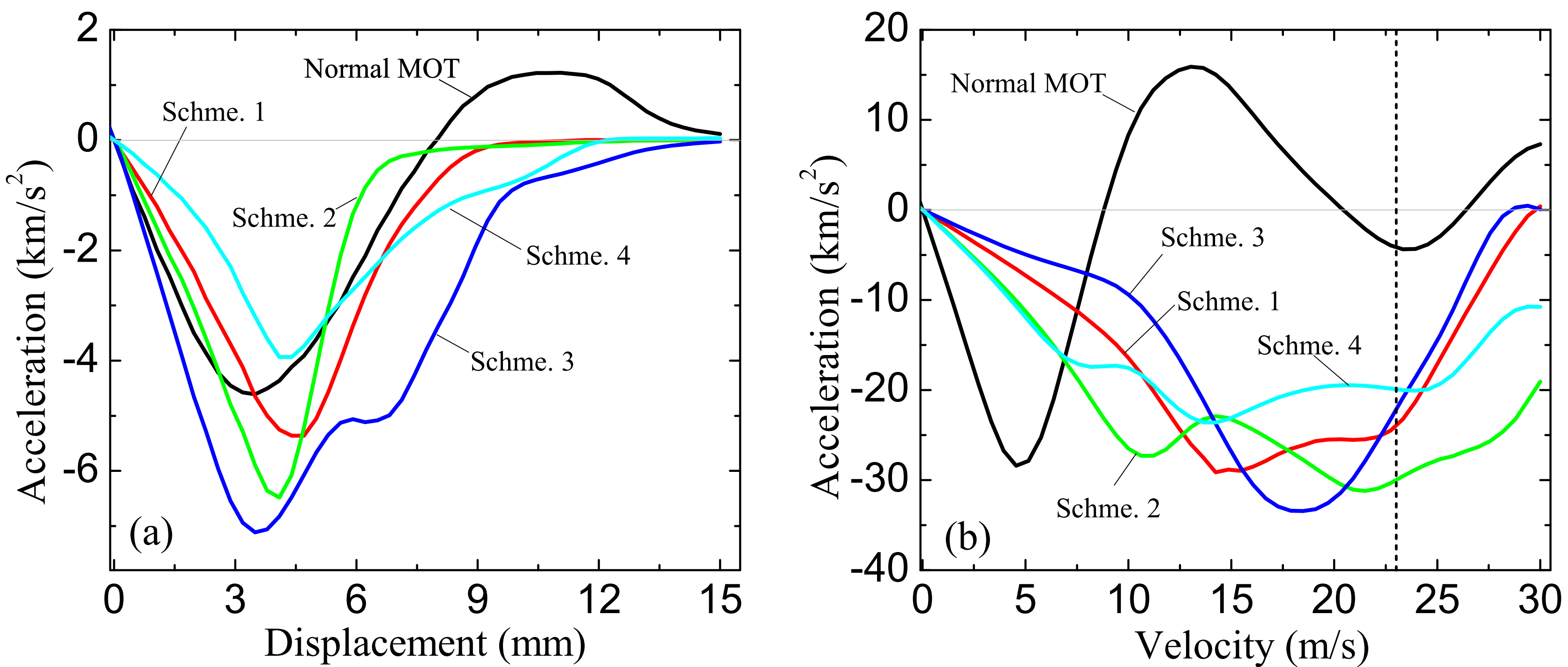}
\caption{Acceleration versus (a) displacement and (b) speed, for four optimized sets of 
parameters. The acceleration of the MOT scheme in [29] is also plotted for comparison. 
The laser power is fixed to $200$\,mW in total. 
}
\label{Fig7}
\end{figure*}

\begin{figure}
\centering
\includegraphics[scale=0.55]{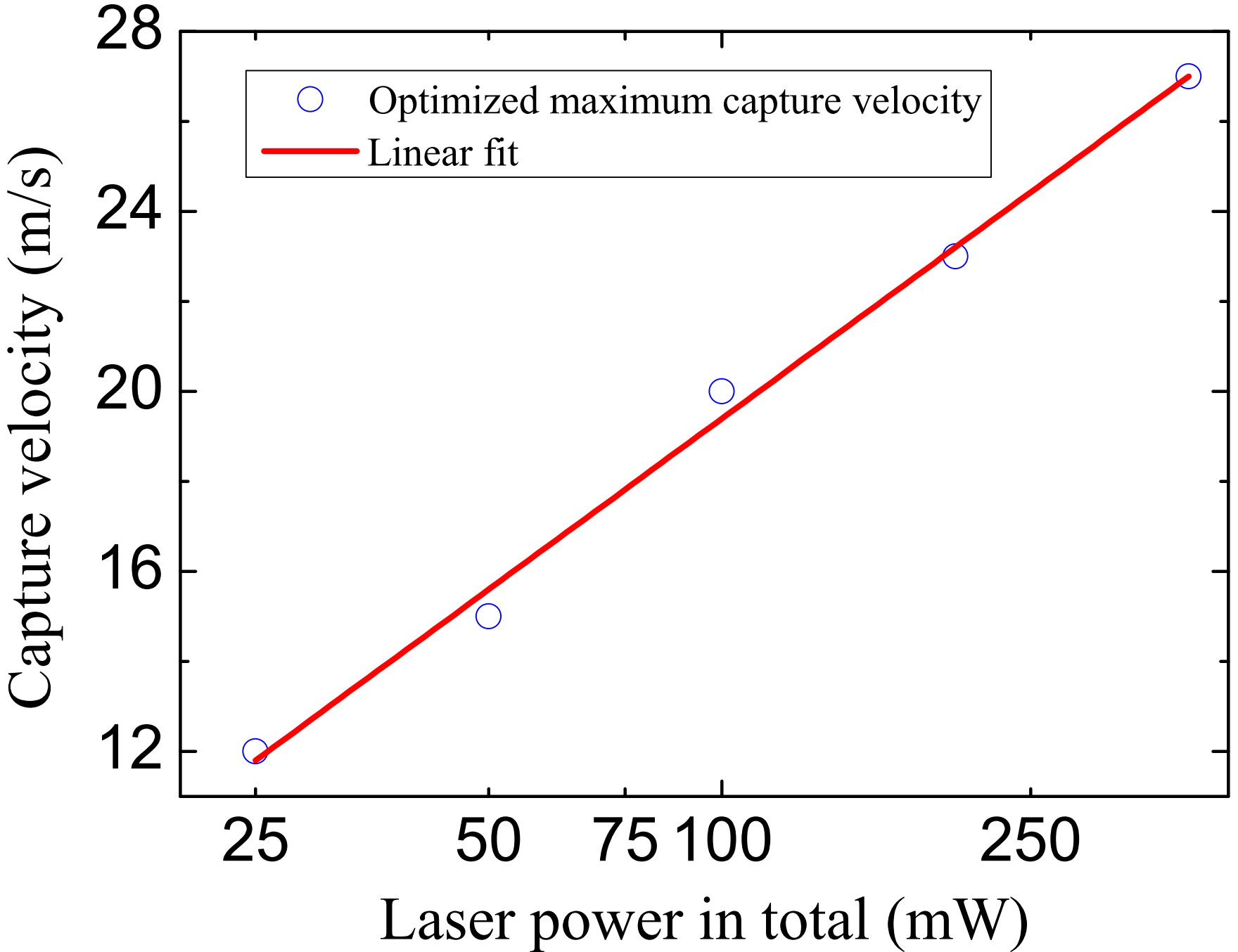}
\caption{Maximum capture velocity versus total laser power for \X $\rightarrow$ \B MOT. The $x$ axis 
is in logarithm scale. 
}
\label{Fig8}
\end{figure}

Fig. \ref{Fig7}(a) and (b) compare the confining force and the cooling force of the conventional \X to \B MOT scheme [29] to the schemes resulting from the optimization.  First of all, it is important to note that the inversion of the forces observed in the conventional MOT scheme disappears for the optimized schemes 1 to 4. Unlike the normal MOT scheme, where the damping acceleration is inverted at 9 m/s, there is strong cooling force for speeds up to 30 m/s for the four new schemes. % shows the position dependent acceleration curve of a molecule, in which the normal MOT means the suggested configuration for \X - \B transition [29], except that $A = 15$\,G/cm and $w=8$\,mm. 
All the optimized schemes have a plateau at large displacement, where the confining force gradually tends to zero. Scheme 3 has the largest peak acceleration of $7120$\,m/s$^2$, and the maximum trap frequency of $243$\,Hz, but its damping coefficient is quite small as shown in Fig.\,\ref{Fig7}(b). A compromise of trapping and cooling is scheme 2, where the trap frequency is $198$\,Hz, and the damping coefficient is $2022$\,s$^{-1}$. The dashed vertical line in Fig.\,\ref{Fig7}(b) reflects the maximum capture velocity, 
$23$\,m/s.

The maximum capture velocity of \X $\rightarrow$  \B MOT versus the total laser power is shown in Fig.\,\ref{Fig8}, in which the blue circles are simulated results and the red solid line is a logarithmic fit to the maximum capture velocity as a function of power. Each data point is the optimal result after 500 iterations. We plot these results here for guiding our future experiments in how much laser power is needed to obtain enough capture velocity. 
Note that the success of this scheme depends on the exact value of a leakage from the \B to the \A state which is currently unknown. Any loss $>10^{-7}$ would complicate a \B-state MOT because the $N=0$ and $N=2$ states would need to be reintroduced into the cooling cycle. 
\section{Conclusion}
In conclusion, we applied the Bayesian optimization approach to search for the maximum capture velocity of a molecular MOT, in which 14 - 17 totally independent parameters are considered. We obtained a group of configurations which can give a 
capture velocity of $24$\,m/s for the \X $\rightarrow$  \A transition at $200$\,mW, along with a large amount of choices with moderate capture velocities. For the \X to \B transition, the BO method also gives possible choices with a large capture velocity. We further studied the maximum capture velocity under different values of laser power for both kinds of transition, and find a logarithmic dependency of the capture velocity on laser power. The laser configurations found through this optimization are experimentally feasible and robust with respect to small changes in parameters. We have shown Bayesian optimization to be a great tool in finding parameters that optimize experiments. Other possible uses for the technique is searching for the best trap frequency or damping coefficient after MOT loading, or to investigate molecule capturing with even less laser 
components. Our approach is completely general and may be used, for example, to investigate possible MOT configurations for the heavier molecules which currently suffer from low capture velocities.

\section*{Acknowledgement}

 P. K., M. St. and M. S. thank the DFG for financial support through RTG 1991. We gratefully acknowledge financial support through  Germany’s Excellence Strategy – EXC-2123/1 QuantumFrontiers. 

\newpage
\appendix

\section{Rotation Matrix}
\label{Appdx:Rot_Mat}
We initially define a group of unit polarization matrices within the lab frame, 
\begin{equation}
\sigma^+=\frac{\sqrt{2}}{2}\times{
\left[ \begin{array}{r}
1 \\
-\textit{i}\\
0 
\end{array} 
\right ]},
\sigma^-=\frac{\sqrt{2}}{2}\times{
\left[ \begin{array}{r}
1 \\
\textit{i} \\
0 
\end{array} 
\right ]},
\Pi=
\left[ \begin{array}{r}
0 \\
0 \\
1 
\end{array} 
\right ],
\end{equation}
where z axis is the quantization axis (QA).

\begin{figure}[htbp]
\centering
\includegraphics[scale=0.2]{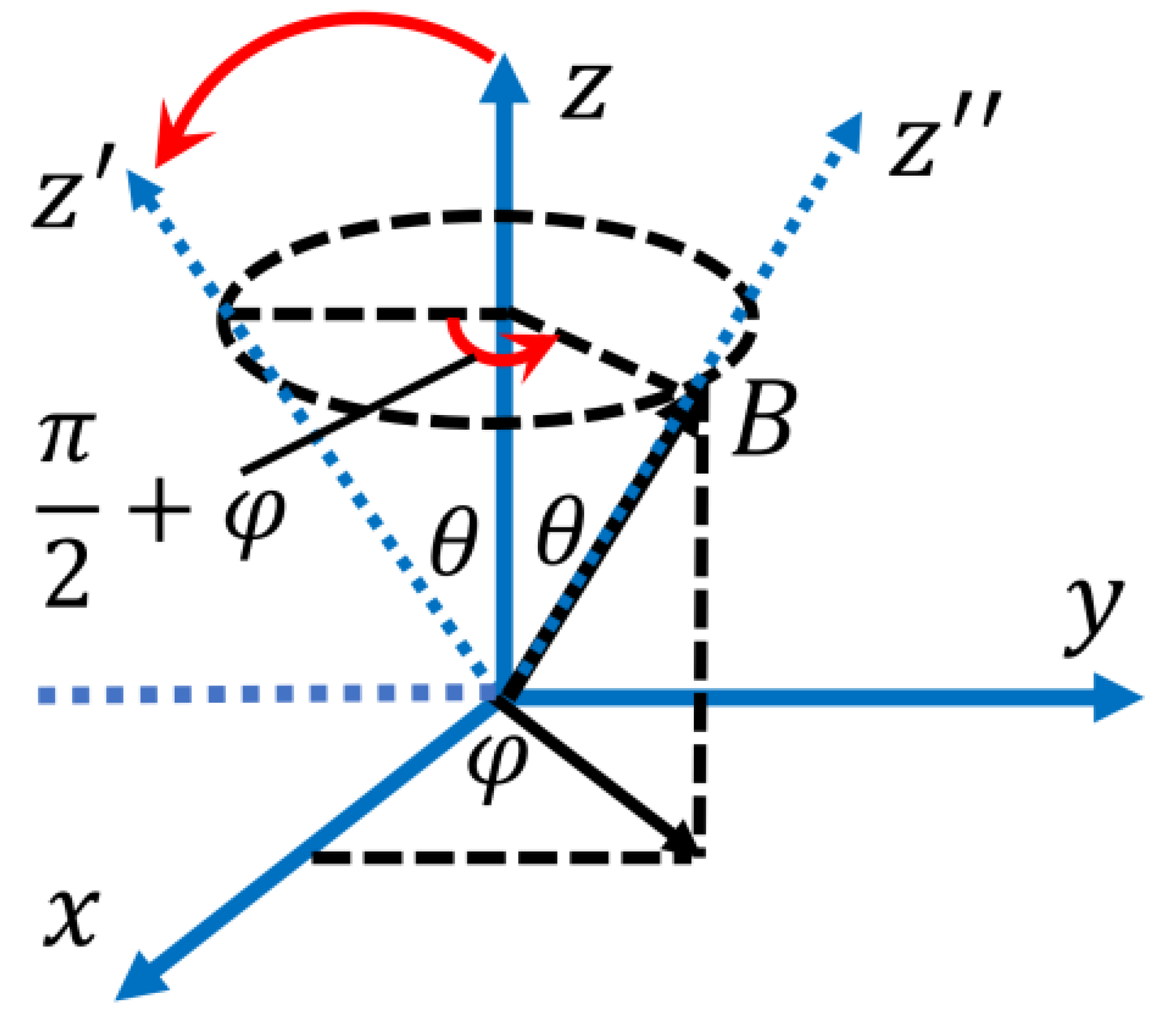}
\caption{Diagram of the rotation process, the $\theta$ is the angle between the magnetic field \textit{B} and the positive \textit{z} axis, the $\varphi$ is the azimuthal angle. The blue solid arrow defines the lab frame, the blue dashed arrow reflects the variation of \textit{z} axis during rotation process and the red arrow reflects the rotational direction. 
}
\label{Fig9}
\end{figure}

Once the magnetic field B is not along the QA direction, we use two-step rotation to get the polarization matrices of $(\sigma^+,\sigma^-,\Pi)$ under the local frame, in which B defines the QA direction, as illustrated in Fig.\,\ref{Fig9}. The first step is rotating about x axis by an angle $\theta$, in which the matrix is 
\begin{equation}
R_x=
\left[ \begin{array}{ccc}
1 & 0 & 0\\
0& \cos(\theta) & -\sin(\theta)\\
0 & \sin(\theta) & \cos(\theta)
\end{array}
\right ]
\end{equation}
The second step is rotating from $z^{\prime}$ to $z^{\prime\prime}$ around $z$ axis with an angle of $\frac{\pi}{2}+\varphi$

\begin{equation}
R_z=
\left[ \begin{array}{ccc}
\cos(\frac{\pi}{2}+\varphi) & -\sin(\frac{\pi}{2}+\varphi) & 0\\
\sin(\frac{\pi}{2}+\varphi)& \cos(\frac{\pi}{2}+\varphi) & 0\\
0 & 0 & 1
\end{array}
\right ]
\end{equation}

Finally, we just project our laser polarization under the lab frame on the three unit matrices under the local frame to get the real transition ratio in $(\sigma^+,\sigma^-,\Pi)$.

\newpage
\section*{References}
\bibliography{bayesian}
\bibliographystyle{iopart-num}

\end{document}